# High-field solution state DNP using cross-correlations

*Maria Grazia Concilio[1,2], Murari Soundararajan[3], Lucio Frydman[1,3]*, Ilya Kuprov[2,*]*

[1]Department of Chemical and Biological Physics, Weizmann Institute of Science, Rehovot, Israel.
[2]School of Chemistry, University of Southampton, Southampton, UK.
[3]National High Magnetic Field Laboratory, Tallahassee, Florida, USA.

[*] *i.kuprov@soton.ac.uk; lucio.frydman@weizmann.ac.il*

**Abbreviations**: DNP, dynamic nuclear polarization; HF, hyperfine; CSA, chemical shift anisotropy; TROSY, transverse relaxation optimised spectroscopy; MW, microwave.

**Highlights**

1. Solution state Overhauser DNP mechanism becomes inefficient at high magnetic fields.
2. Numerical simulations indicate that high-field DNP still apparently exists in some systems.
3. A detailed theoretical analysis finds multiple cross-correlated (CC) relaxation processes.
4. CCDNP effect remains efficient at high magnetic fields with moderate MW power.

**Graphical abstract**:

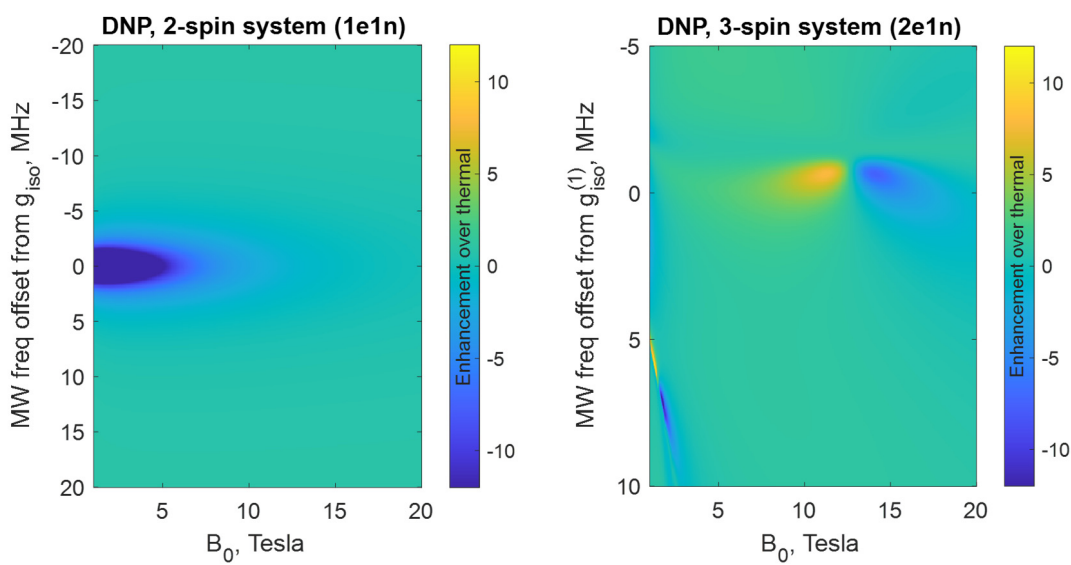

## Abstract

At the magnetic fields of common NMR instruments, electron Zeeman frequencies are too high for efficient electron-nuclear dipolar cross-relaxation to occur in solution. The rate of that process fades with the electron Zeeman frequency as $\omega^{-2}$ – in the absence of isotropic hyperfine couplings, liquid state dynamic nuclear polarisation (DNP) in high-field magnets is therefore impractical. However, contact coupling and dipolar cross-relaxation are not the only mechanisms that can move electron magnetisation to nuclei in liquids: multiple cross-correlated (CC) relaxation processes also exist, involving various combinations of interaction tensor anisotropies. The rates of some of those processes have more favourable high-field behaviour than dipolar cross-relaxation, but due to the difficulty of their numerical – and particularly analytical – treatment, they remain largely uncharted. In this communication, we report analytical evaluation of every rotationally driven relaxation process in liquid state for 1e1n and 2e1n spin systems, as well as numerical optimisations of the steady-state DNP with respect to spin Hamiltonian parameters. A previously unreported cross-correlation DNP (CCDNP) mechanism was identified for the 2e1n system, involving multiple relaxation interference effects and inter-electron exchange coupling. Using simulations, we found realistic spin Hamiltonian parameters that yield stronger nuclear polarisation at high magnetic fields than dipolar cross-relaxation.

## 1. Introduction

Dynamic nuclear polarization (DNP) was discovered [1,2] shortly after magnetic resonance itself, and recently saw a renaissance, particularly in solid state [3] and in medical imaging [4], with much work published on the instruments [5-8], radicals [9-12], and theory [13-16] of the process.

One unsolved problem is achieving strong liquid state DNP in high-field magnets (~7 Tesla and above) that dominate modern NMR facilities. At those fields, electron-nuclear dipolar cross-relaxation is too slow in common solvents [17-19]. This problem appears because the spectral power densities in the electron-nuclear cross-relaxation rate $\sigma_{E,N}$ drop as the square of the electron Zeeman frequency:

$$\sigma_{E,N} \approx \frac{\gamma_E^2 \gamma_N^2 \hbar^2}{10} \left(\frac{\mu_0}{4\pi}\right)^2 \frac{\tau_C}{r_{EN}^6} \left( \frac{6}{1+(\omega_E+\omega_N)^2 \tau_C^2} - \frac{1}{1+(\omega_E-\omega_N)^2 \tau_C^2} \right) \quad (1)$$

Here, E and N indices refer to the electron and the nucleus, $\gamma$ are magnetogyric ratios, $\tau_C$ is the rotational correlation time, $r_{EN}$ is the electron-nucleus distance, and $\omega = -\gamma B_0$ are the Zeeman frequencies in the external magnetic field $B_0$. For common electron-nucleus pairs in non-viscous solvents and high-field magnets, $\tau_C \sim 10$ ps and $\omega_E/2\pi \sim 100$ GHz, meaning that $(\omega_E \pm \omega_N)^2 \tau_C^2 \gg 1$ and resulting in $\sigma_{E,N}$ below 1 Hz – too slow to build up useful levels of nuclear spin polarisation.

The DNP community had seen a high-field bottleneck before – in solid state DNP, where two-spin transitions with unfavourable magnetic field dependence ("solid effect") were initially used [20]. It was later found that a better approach was to use a three-spin process ("cross effect") whose rate has a more favourable magnetic field scaling [13,21]. That bottleneck was in the coherent Hamiltonian; the one we consider here is in the relaxation superoperator.

It occurred to us that the relaxation-driven dynamics responsible for liquid state DNP could also have significant three-spin effects. The high dimension of the three-spin Liouville space and the abundance of uncharted relaxation processes makes the matter worth looking at: a rigid molecule with three anisotropic Zeeman interactions, two anisotropic hyperfine interactions, and an inter-electron dipolar coupling has 15 different cross-correlations [22]. Under the right conditions, the electron magnetisation could potentially reach the nucleus through these cross-correlated (CC) relaxation channels.

In this communication, we report theoretical evidence that cross-correlated DNP (CCDNP) exists and can produce significant nuclear polarisation well above the fields (~3 Tesla) at which electron-nuclear dipolar cross-relaxation becomes inactive in liquids. To identify this process, we have deployed a brute-force numerical simulation tool [23-25] that accounts for every cross-correlation within Bloch-Redfield-Wangsness (BRW) relaxation theory [26,27], as well as a symbolic processing engine [28] that cuts through the voluminous perturbation theory treatment and returns analytical expressions for the various elements of BRW relaxation superoperator. A detailed analysis of system trajectories yielded the mechanism which we present here. CCDNP does not require isotropic hyperfine couplings and remains effective in the presence of inter-electron exchange interaction.

## 2. Theoretical methods

Laboratory frame spin Hamiltonians with continuous microwave irradiation at a fixed frequency $\omega_{\text{MW}}$ were generated using *Spinach* library [24]:

$$\hat{H} = \hat{\vec{S}}^{(\text{E1})} \cdot \mathbf{Z}^{(\text{E1})} \cdot \vec{B}_0 + \hat{\vec{S}}^{(\text{E2})} \cdot \mathbf{Z}^{(\text{E2})} \cdot \vec{B}_0 + \hat{\vec{S}}^{(\text{N})} \cdot \mathbf{Z}^{(\text{N})} \cdot \vec{B}_0 +$$
$$+ \hat{\vec{S}}^{(\text{E1})} \cdot (\mathbf{D} + J) \cdot \hat{\vec{S}}^{(\text{E2})} + \hat{\vec{S}}^{(\text{E1})} \cdot \mathbf{A}^{(1)} \cdot \hat{\vec{S}}^{(\text{N})} + \hat{\vec{S}}^{(\text{E2})} \cdot \mathbf{A}^{(2)} \cdot \hat{\vec{S}}^{(\text{N})} + \quad (2)$$
$$+ \frac{B_1 \cos(\omega_{\text{MW}} t)}{3} \left( \text{Tr}\left[\mathbf{Z}^{(\text{E1})}\right] \hat{S}_{\text{X}}^{(\text{E1})} + \text{Tr}\left[\mathbf{Z}^{(\text{E2})}\right] \hat{S}_{\text{X}}^{(\text{E2})} \right)$$

where $\mathbf{Z}^{(k)}$ are Zeeman tensors of the indicated particles, $\mathbf{D}$ is the inter-electron dipolar interaction tensor in the point magnetic dipole approximation, $J$ is the inter-electron scalar (*aka* "exchange") coupling in angular frequency units, $\mathbf{A}^{(k)}$ are the hyperfine interaction tensors of the nucleus with the indicated electrons, $\vec{B}_0$ is the external magnetic field, $B_1$ is the magnetic field associated with the microwave irradiation (assumed to be directed along the X axis of the laboratory frame), and $\hat{\vec{S}}^{(k)} = \begin{bmatrix} \hat{S}_{\text{X}}^{(k)} & \hat{S}_{\text{Y}}^{(k)} & \hat{S}_{\text{Z}}^{(k)} \end{bmatrix}^{\text{T}}$ are spin operator vectors associated with the indicated particles.

Brute-force analytical [28] and numerical [23,25] implementations of Bloch-Redfield-Wangsness relaxation theory [26,27] were used to obtain relaxation superoperators. Because the systems in question are at room temperature, inhomogeneous thermalisation [29] was used in the master equation, wherein the product of the density matrix and the relaxation superoperator $\hat{\hat{R}}\hat{\rho}$ is heuristically replaced by $\hat{\hat{R}}(\hat{\rho} - \hat{\rho}_{\text{eq}})$. Under liquid state DNP conditions, this is a good approximation.

### 2.1 Rotating frame simulation path

A numerical interaction representation transformation [23] was performed with respect to the electron Zeeman Hamiltonian matched to the microwave irradiation frequency:

$$\hat{H}_0 = \omega_{\text{MW}} \left[ \hat{S}_{\text{Z}}^{(\text{E1})} + \hat{S}_{\text{Z}}^{(\text{E2})} \right] \quad (3)$$

To first order in the average Hamiltonian theory [30], this transformation makes the microwave irradiation part of the Hamiltonian time-independent, while retaining secular parts of the inter-electron couplings, as well as secular and pseudosecular parts of the electron-nuclear couplings, in the interaction part [31]. In Liouville space, the equation of motion becomes:

$$\frac{\partial \hat{\rho}}{\partial t} = -i\hat{\hat{H}}\hat{\rho} + \hat{\hat{R}}(\hat{\rho} - \hat{\rho}_{\text{eq}}) \quad (4)$$

where the double hat on the Hamiltonian indicates a commutation superoperator, $\hat{\hat{R}}$ is the symmetric (not corrected for thermal equilibrium) negative definite relaxation superoperator, $\hat{\rho}$ is the density matrix and the 'eq' subscript indicates thermal equilibrium. At the steady state ($t = \infty$), the time derivative is zero, and the steady state density matrix $\hat{\rho}_\infty$, at the point where microwave irradiation is balanced out by relaxation, is obtained as:

$$-i\hat{\hat{H}}\hat{\rho}_\infty + \hat{\hat{R}}(\hat{\rho}_\infty - \hat{\rho}_{\text{eq}}) = 0 \quad \Rightarrow \quad \hat{\rho}_\infty = \left(\hat{\hat{R}} - i\hat{\hat{H}}\right)^{-1} \hat{\hat{R}}\hat{\rho}_{\text{eq}} \quad (5)$$

An important feature here is the matrix-inverse-times-vector operation, which is numerically cheaper than the matrix inverse. The steady state nuclear magnetisation $\langle N_Z \rangle_\infty$ is then obtained from $\hat{\rho}_\infty$:

$$\left\langle S_Z^{(N)} \right\rangle_\infty = \text{Tr}\left[ \hat{\rho}_\infty \hat{S}_Z^{(N)} \right] \tag{6}$$

This quantity was calculated as a function of external magnetic field, rotational correlation time, and microwave frequency offset from the isotropic Zeeman frequency of one of the electrons. The offset variation is necessary because maximum DNP does not necessarily correspond to the microwave irradiation being exactly on resonance, much like it happens in the solid state [13,32].

## 2.2 Laboratory frame simulation path

It is possible to avoid the rotating frame transformation and run the same calculation in the laboratory frame using the Fokker-Planck formalism [33], which handles the time dependence in the microwave term using the fact that the exponential of a derivative operator is a finite shift operator:

$$\exp\left[ \omega t \frac{\partial}{\partial \varphi} \right] f(\varphi) = f(\varphi + \omega t) \tag{7}$$

The Fokker-Planck formalism views the density matrix as a function of both the microwave phase $\varphi$ and time $t$. For periodic processes, this may be seen as a reformulation of Floquet theory [34,35]. In our current DNP context, the equation of motion is:

$$\frac{\partial}{\partial t} \hat{\rho}(\varphi,t) = \underbrace{-i\hat{\hat{H}}(\varphi)\hat{\rho}(\varphi,t)}_{\text{coherent dynamics}} + \underbrace{\hat{\hat{R}}\left(\hat{\rho}(\varphi,t) - \hat{\rho}_{eq}\right)}_{\text{relaxation}} + \underbrace{\omega_{MW} \frac{\partial}{\partial \varphi} \hat{\rho}(\varphi,t)}_{\text{MW phase turning}} \tag{8}$$

where $\varphi$ is the microwave phase that now replaces the $\omega_{MW} t$ term in Equation (2). At the steady state ($t = \infty$), the time derivative is zero:

$$-i\hat{\hat{H}}(\varphi)\hat{\rho}_\infty(\varphi) + \hat{\hat{R}}\left(\hat{\rho}_\infty(\varphi) - \hat{\rho}_{eq}\right) + \omega_{MW} \frac{\partial}{\partial \varphi} \hat{\rho}_\infty(\varphi) = 0 \tag{9}$$

This is now an ordinary differential equation for the steady state orbit $\hat{\rho}_\infty(\varphi)$ that can be solved numerically on a finite grid over the microwave phase angle. On an *N*-point periodic grid $\{\varphi_k\}$, the Hamiltonian, the density matrix, and the relaxation superoperator are block-replicated:

$$\hat{\hat{H}}(\varphi) \rightarrow \begin{pmatrix} \hat{\hat{H}}_1 & & \\ & \cdots & \\ & & \hat{\hat{H}}_N \end{pmatrix}, \qquad \hat{\hat{H}}_k = \hat{\hat{H}}(\varphi_k)$$

$$\hat{\hat{R}} \rightarrow \begin{pmatrix} \hat{\hat{R}} & & \\ & \cdots & \\ & & \hat{\hat{R}} \end{pmatrix}, \qquad \hat{\rho}(\varphi) \rightarrow \begin{pmatrix} \hat{\rho}_1 \\ \vdots \\ \hat{\rho}_N \end{pmatrix}, \qquad \hat{\rho}_k = \hat{\rho}_\infty(\varphi_k) \tag{10}$$

and the differential operator $\partial/\partial \varphi$ becomes a matrix – the best choice for periodic dynamics is a Fourier differentiation matrix [36], which we will denote $\partial^N$. With all of that in place, Eq (9) becomes:

$$-i\begin{pmatrix}\hat{\hat{H}}_1 & & \\ & \cdots & \\ & & \hat{\hat{H}}_N\end{pmatrix}\begin{pmatrix}\hat{\rho}_1 \\ \vdots \\ \hat{\rho}_N\end{pmatrix}+\begin{pmatrix}\hat{\hat{R}} & & \\ & \cdots & \\ & & \hat{\hat{R}}\end{pmatrix}\begin{pmatrix}\hat{\rho}_1-\hat{\rho}_{eq} \\ \vdots \\ \hat{\rho}_N-\hat{\rho}_{eq}\end{pmatrix}+\omega_{MW}\left[\partial^N\otimes\mathbf{1}_{spin}\right]\begin{pmatrix}\hat{\rho}_1 \\ \vdots \\ \hat{\rho}_N\end{pmatrix}=0 \quad (11)$$

This is now an algebraic equation for the steady state orbit; it was solved in *Matlab* using the ILU-preconditioned [37] GMRES method [38]:

$$\begin{pmatrix}\hat{\rho}_1 \\ \vdots \\ \hat{\rho}_N\end{pmatrix}=\left[-i\begin{pmatrix}\hat{\hat{H}}_1+i\hat{\hat{R}} & & \\ & \cdots & \\ & & \hat{\hat{H}}_N+i\hat{\hat{R}}\end{pmatrix}+\omega_{MW}\left[\partial^N\otimes\mathbf{1}_{spin}\right]\right]^{-1}\begin{pmatrix}\hat{\hat{R}}\hat{\rho}_{eq} \\ \vdots \\ \hat{\hat{R}}\hat{\rho}_{eq}\end{pmatrix} \quad (12)$$

The steady state nuclear magnetisation was obtained by averaging over the phase grid and taking a scalar product with the corresponding operator:

$$\left\langle S_Z^{(N)}\right\rangle_\infty=\left\langle \hat{S}_Z^{(N)}\left|\frac{1}{N}\sum_{k=1}^N\hat{\rho}_k\right.\right\rangle \quad (13)$$

For the calculations reported in this work, the laboratory frame and the rotating frame implementations produced identical answers. Independent implementations using two dissimilar formalisms were necessary to ensure reliability, in view of the purely theoretical nature of this study.

## 3. One electron and one nucleus

In a system with a spin-1/2 electron and a spin-1/2 nucleus, three anisotropic interactions cause relaxation when a rigid molecule undergoes rotational diffusion in solution: the electron Zeeman interaction (*g*-tensor anisotropy, G), the nuclear Zeeman interaction (chemical shift anisotropy, CSA), and the electron-nuclear hyperfine (HF) coupling. These lead to three cross-correlations: G-CSA between the anisotropies of the two Zeeman tensors, HF-CSA between the hyperfine coupling and the chemical shift anisotropy, and HF-G between the hyperfine coupling and the *g*-tensor anisotropy. Some of these mechanisms have been seen in nuclear polarisation enhancement processes before [39,40].

A brute-force numerical simulation for a system designed to have significant G-CSA, HF-CSA, and HF-G cross-correlations shows an intriguing increase in the steady state nuclear magnetisation in high field (Figure 1A). This increase disappears if either the *g*-tensor anisotropy or the chemical shift anisotropy are zeroed, and also if the isotropic hyperfine coupling is removed (Figure 1, other panels). This behaviour is typical of cross-correlated relaxation [39,40], but the longitudinal process:

$$\hat{E}_Z \xrightarrow{\text{HF-G}} 2\hat{E}_Z\hat{N}_Z \xrightarrow{\text{HF-CSA}} \hat{N}_Z \quad (14)$$

cannot be the explanation because the high-field bottleneck is still present – the rate of the first stage has the square of the electron Zeeman frequency in the denominator:

$$R\left[\hat{E}_Z\to 2\hat{E}_Z\hat{N}_Z\right]=-\frac{2}{15}\aleph_{HF,G}J(\omega_E) \qquad J(\omega)=\frac{\tau_C}{1+\tau_C^2\omega^2} \quad (15)$$
$$R\left[2\hat{E}_Z\hat{N}_Z\to\hat{N}_Z\right]=-\frac{2}{15}\aleph_{HF,CSA}J(\omega_N)$$

where the second rank scalar product $\aleph_{\mathbf{A,B}}$ of two 3×3 interaction tensors $\mathbf{A}$ and $\mathbf{B}$ is obtained by polarising the second-rank norm [41]:

$$\aleph_{\mathbf{A,B}} = \frac{\Delta^2_{\mathbf{A+B}} - \Delta^2_{\mathbf{A-B}}}{4}$$

$$\Delta^2_{\mathbf{A}} = a^2_{XX} + a^2_{YY} + a^2_{ZZ} - a_{XX}a_{YY} - a_{XX}a_{ZZ} - a_{YY}a_{ZZ} +$$
$$+ \frac{3}{4}\left[(a_{XY} + a_{YX})^2 + (a_{XZ} + a_{ZX})^2 + (a_{YZ} + a_{ZY})^2\right]$$ (16)

The fact that DNP enhancement is nonetheless present in Figure 1A indicates that the bottleneck stage in Equations (14) and (15) is somehow being bypassed.

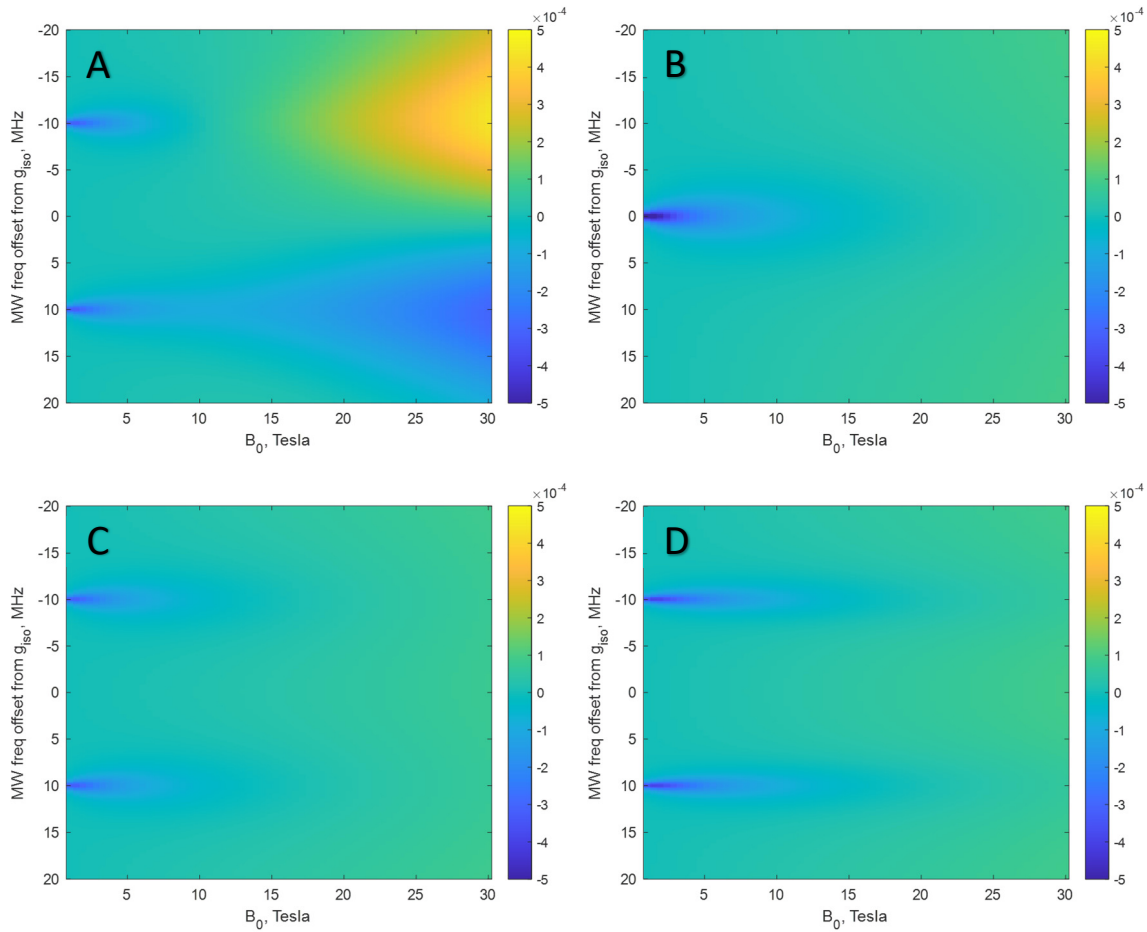

**Figure 1.** *Absolute proton magnetisation at the steady state for a proton-electron system with the following parameters: chemical shift tensor eigenvalues [15 5 –20] ppm, g-tensor eigenvalues [2.00210 2.00250 2.00290], ZYZ Euler angles of the g-tensor relative to the chemical shift tensor [π/3 π/4 π/5], electron-nuclear distance 3 Angstrom along Z axis, electron-nuclear isotropic hyperfine coupling 20 MHz, rotational correlation time $10^{-11}$ seconds. **(A):** Spin system as described. **(B):** As described but with isotropic hyperfine coupling set to zero. **(C):** As described but with nuclear chemical shift anisotropy set to zero. **(D):** As described but with electron g-tensor anisotropy set to zero. Simulation scripts are available in the example set of Spinach version 2.6 and later.*

A detailed inspection of the system trajectory reveals that the bottleneck is bypassed via a combination of microwave irradiation and transverse HF-G cross-correlation [39]:

$$\hat{E}_Z \xrightarrow{\text{MW}} \hat{E}_\pm \xrightarrow{\text{HF-G}} 2\hat{E}_\pm\hat{N}_Z \xrightarrow{\text{MW}} 2\hat{E}_Z\hat{N}_Z \xrightarrow{\text{HF-CSA}} \hat{N}_Z$$ (17)

This process is amplified at high field because the rate of transverse HF-G cross-correlation:

$$R\left[\hat{E}_{\pm} \to 2\hat{E}_{\pm}\hat{N}_{Z}\right] = -\frac{\aleph_{\text{HF,G}}}{45}\left[4J(0) + 3J(\omega_{\text{E}})\right] \tag{18}$$

increases with magnetic field due to the presence of the electron Zeeman interaction in combination with the field-independent $J(0)$ term in the spectral power density. There are also other, less prominent and more complicated, combinations of microwave and cross-correlation processes that link electron and nuclear magnetisation in this system. A *Mathematica* worksheet containing a complete relaxation analysis may be found in the Supplementary Information.

Strong high-field DNP disappears if either of the two Zeeman anisotropies is set to zero (Figures 1C, 1D), indicating that a cross-correlation pathway is indeed responsible. In common with phenomena like TROSY [42], the effect also disappears (Figure 1B) in the absence of isotropic coupling: for any differential relaxation process to be feasible, some multiplet structure must be present [43]. This last requirement is a critical weakness of this two-spin mechanism: when isotropic hyperfine couplings are present conventional DNP already works well at high fields [44], and better ways of transferring electron magnetisation to nuclei already exist using coherent dynamics [45].

This suggests that two-electron systems would be more promising because the nearly unavoidable isotropic *inter-electron* coupling may be used to take advantage of TROSY type effects in the two-electron subspace. Accordingly, we turn our attention to DNP in 2e1n three-spin system.

## 4. Two electrons and one nucleus

The large number of anisotropic interactions and the large dimension of the state space of the 2e1n spin system necessitates a brute-force numerical search for parameter combinations that show rapid electron-nuclear polarisation transfer in high fields. We used this approach to identify promising parameter combinations, and then ran detailed trajectory analysis to determine the mechanism.

### 4.1 Parameter space search

To obtain promising parameter combinations for detailed relaxation analysis, we started from educated guesses involving spin-1/2 biradicals containing trityl [46] and oxochromate [47] spin centres, and performed brute-force numerical optimisations of simulated steady-state nuclear magnetisation with respect to all chemically controllable parameters of the spin Hamiltonian. This is a useful strategy because *Spinach* [24] contains a formally complete implementation of BRW relaxation theory [23,25], and may be relied upon to pick up unusual effects [48,49].

Steady state nuclear magnetisation was optimised as a function of spin system parameters using Gaussian processes machine learning [50], as implemented in the M-Loop package [51]. The following target functional was used:

$$\Omega(\mathbf{x}) = \max\left\{-\min_{\omega_{\text{MW}}}\left\langle N_{Z}(\mathbf{x},\omega_{\text{MW}})\right\rangle_{\infty}, \max_{\omega_{\text{MW}}}\left\langle N_{Z}(\mathbf{x},\omega_{\text{MW}})\right\rangle_{\infty}\right\} \tag{19}$$

which dictates that, across the microwave frequency sweep range, either the largest positive nuclear polarisation be further maximised, or the smallest negative nuclear polarisation be further minimised with respect to the spin system parameter array $\mathbf{x}$.

*Table 1. Parameters of the spin system used for the in-depth relaxation mechanism analysis in this work. This system was found, by a brute-force numerical optimisation of simulated nuclear magnetisation, to have a much larger steady-state DNP amplitude at high field than could be expected from direct electron-nuclear dipolar cross-relaxation.*

| Parameter | Value |
| --- | --- |
| $^1$H chemical shift tensor eigenvalues, [xx yy zz], ppm | [0 10 20] |
| $^1$H chemical shift tensor, ZYZ active Euler angles, rad | [0 0 0] |
| Electron 1 *g*-tensor eigenvalues, [xx yy zz] | [2.0034 2.0038 2.0038] |
| Electron 1 *g*-tensor, ZYZ active Euler angles, rad | [−0.872 −0.013 0.868] |
| Electron 2 *g*-tensor eigenvalues, [xx yy zz] | [2.0057 2.0030 2.0030] |
| Electron 2 *g*-tensor, ZYZ active Euler angles, rad | [−1.145 0.061 1.143] |
| $^1$H coordinates, [x y z], Angstrom | [0 0 0] |
| Electron 1 coordinates, [x y z], Angstrom | [5.090 0.010 0.958] |
| Electron 2 coordinates, [x y z], Angstrom | [−5.090 0.061 1.032] |
| Rotational correlation time, ps | 100 |
| Electron nutation frequency under MW irradiation, MHz | 1.0 |
| Electron-electron exchange coupling, MHz | 3.0 |

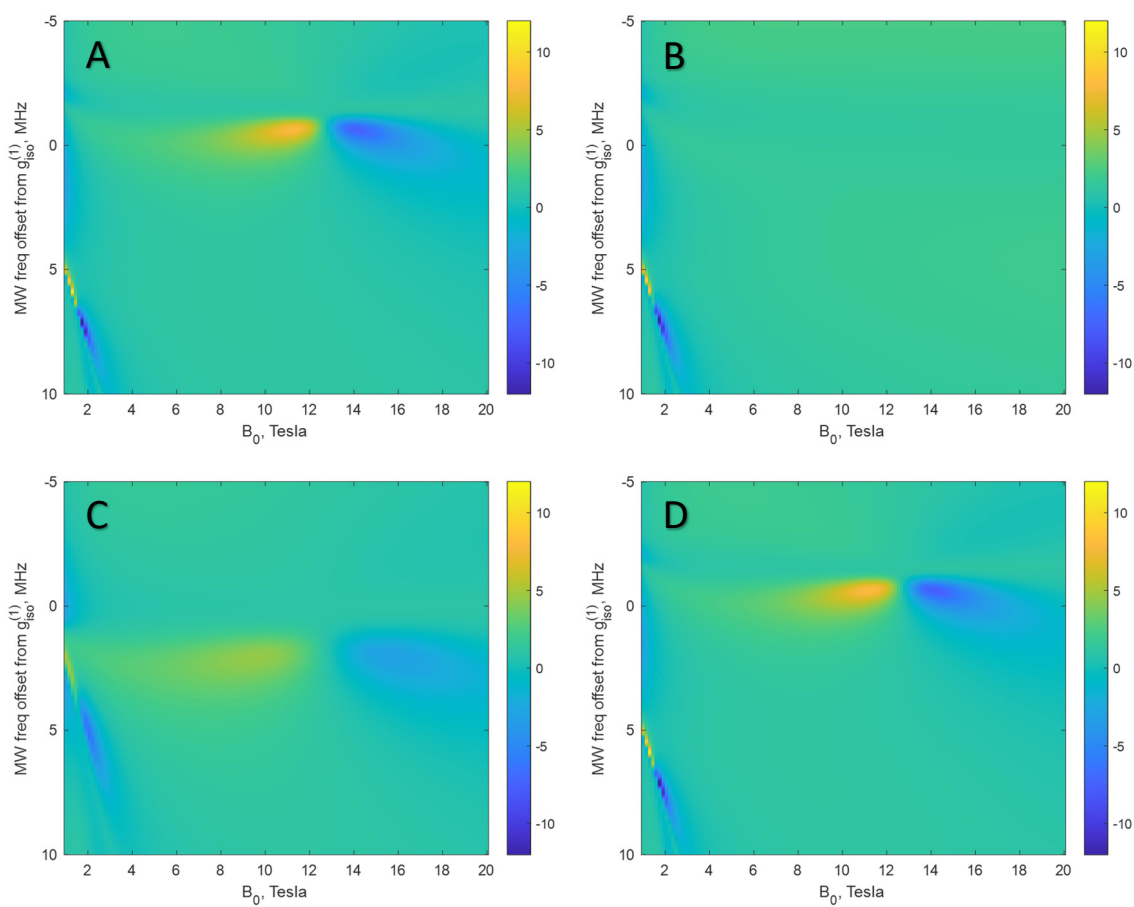

*Figure 2. Steady state longitudinal nuclear magnetisation (normalised to the thermal equilibrium value) as a function of the static magnetic field and the microwave offset relative to the Zeeman frequency of electron 1. (A): Parameters as specified in Table 1. (B): Table 1 with the g-tensor anisotropy on electron 1 set to zero; note how this eliminates the high field nuclear polarisation transfer. (C): Table 1 with the inter-electron exchange coupling set to zero. (D): Table 1 with nuclear chemical shift anisotropy set to zero. Simulation scripts are available in the example set of version 2.6 and later of Spinach library.*

One realistic parameter combination that the optimisation turned up is given in Table 1; there are two more such combinations in Table S1 of the Supplementary Information. Note the absence of isotropic hyperfine interactions in all of these cases: to make CCDNP broadly applicable and to bypass the spin diffusion barrier, we aimed for purely dipolar (*i.e.* zero spherical average in liquids) electron-nuclear interactions. Figure 2A shows the simulated nuclear magnetisation enhancement for the system described in Table 1, as a function of the external magnetic field and the microwave frequency offset; Figure S1 in the SI shows the same for the systems described in Table S1.

### 4.2 Cross-correlated DNP mechanism in biradicals

The steady state nuclear polarisation in the test system (Table 1) shows complicated field and offset dependence (Figure 2) that cannot be explained by electron-nuclear dipolar cross-relaxation. In common with the 1e1n system, it is too slow – at high magnetic fields, its rate decays quadratically as a function of the electron Zeeman frequency:

$$R\left[\hat{E}_Z^{(1)} \to \hat{N}_Z\right] = -\frac{\Delta_{\text{HF1}}^2}{18} J(\omega_{E1}) = -\frac{\Delta_{\text{HF1}}^2}{18}\frac{\tau_C}{1+\omega_E^2\tau_C^2} \approx -\frac{\Delta_{\text{HF1}}^2}{18\omega_E^2\tau_C} \qquad (20)$$

Here and below, we have ignored the nuclear Zeeman frequency when it appears alongside the electron frequency, and neglected the dynamic frequency shifts within BRW theory:

$$J(\omega_E \pm \omega_N) \approx J(\omega_E), \qquad J(-\omega) \approx J(\omega) \qquad (21)$$

The *Mathematica* script performing the analytical relaxation theory analysis [28] may be found in the Supplementary Information. The rates arising from it for the system described in Table 1 are given in Table 2. Clearly, dipolar cross-relaxation is too slow to be of relevance at the $B_0$ fields of interest.

*Table 2.* Rates of electron longitudinal relaxation ($R_{1E}$), nuclear longitudinal relaxation ($R_{1N}$), and electron-nuclear cross-relaxation (σ) at 14.1 Tesla for the system described in Table 1.

| Process | $R_{1E}$, MHz | σ, Hz | $R_{1N}$, Hz |
|---|---|---|---|
| $\hat{E}_Z^{(1)} \to \hat{N}_Z$ | 0.96 | –0.0104 | 667 |
| $\hat{E}_Z^{(2)} \to \hat{N}_Z$ | 0.97 | –0.0104 | |

Other longitudinal cross-relaxation routes out of the $\{\hat{E}_Z^{(1)}, \hat{E}_Z^{(2)}, 2\hat{E}_Z^{(1)}\hat{E}_Z^{(2)}\}$ subspace are also blocked at high field by the asymptotic $\sim 1/\omega_E^2$ Zeeman frequency dependence of their rates:

$$R\left[\hat{E}_Z^{(1)} \to 2\hat{E}_Z^{(1)}\hat{N}_Z\right] = -\frac{2}{15}\aleph_{\text{G1,HF1}} J(\omega_{E1})$$
$$R\left[\hat{E}_Z^{(1)} \to 4\hat{E}_Z^{(1)}\hat{E}_Z^{(2)}\hat{N}_Z\right] = -\frac{1}{15}\aleph_{\text{DD,HF1}} J(\omega_{E1}) \qquad (21)$$

where G1 refers to the *g*-tensor of the first electron, HF1 to the hyperfine coupling between the nucleus and electron 1, and DD to the inter-electron dipolar interaction; the expressions are similar for the other electron. The longitudinal two-electron correlation has a similar problem – its connections to nuclear polarisation states are too slow:

$$R\left[2\hat{E}_Z^{(1)}\hat{E}_Z^{(2)} \to \hat{N}_Z\right] = 0$$

$$R\left[2\hat{E}_Z^{(1)}\hat{E}_Z^{(2)} \to 2\hat{E}_Z^{(1)}\hat{N}_Z\right] = -\frac{\Delta_{HF2}^2}{18}J(\omega_{E2}) - \frac{\aleph_{DD,HF1}}{15}J(\omega_{E1})$$

$$R\left[2\hat{E}_Z^{(1)}\hat{E}_Z^{(2)} \to 2\hat{E}_Z^{(2)}\hat{N}_Z\right] = -\frac{\Delta_{HF1}^2}{18}J(\omega_{E1}) - \frac{\aleph_{DD,HF2}}{15}J(\omega_{E2}) \quad (22)$$

$$R\left[2\hat{E}_Z^{(1)}\hat{E}_Z^{(2)} \to 4\hat{E}_Z^{(1)}\hat{E}_Z^{(2)}\hat{N}_Z\right] = -\frac{2}{15}\aleph_{G1,HF1}J(\omega_{E1}) - \frac{2}{15}\aleph_{G2,HF2}J(\omega_{E2})$$

The only efficient longitudinal process is the exchange of magnetisation between the two electrons when their isotropic Zeeman frequencies are sufficiently close:

$$R\left[\hat{E}_Z^{(1)} \to \hat{E}_Z^{(2)}\right] = \frac{\Delta_{DD}^2}{90}\left[J(\omega_{E1}-\omega_{E2}) - 6J(\omega_{E1}+\omega_{E2})\right]. \quad (23)$$

The fact that significant nuclear polarisation enhancement is nonetheless predicted by the brute-force numerical simulation in Figure 2 indicates that viable transverse routes exist. The effect disappears when *g*-tensor anisotropy is set to zero (Figure 2B); this suggests that at least one of its stages involves a cross-correlated transverse relaxation effect.

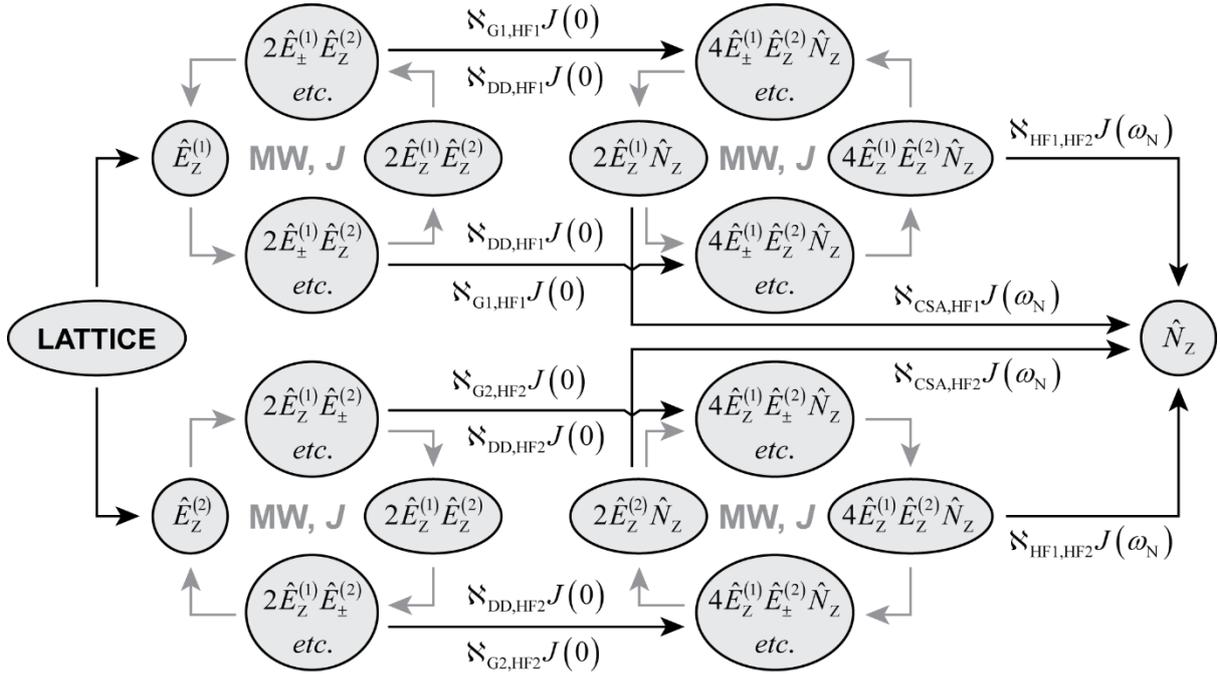

*Figure 3.* CCDNP mechanism in a system with two electrons and a nucleus. Grey arrows indicate coherent dynamics, black arrows indicate relaxation-driven dynamics. At the nominal first stage (at the steady state, all stages are simultaneously active), the microwave Hamiltonian rotates the initial electron longitudinal magnetisation to the transverse magnetisation associated with one of the four lines in the liquid state ESR spectrum. In a well-chosen spin system, the cross-correlation between *g*-tensor anisotropy and inter-electron dipolar interaction makes this state relatively long-lived (Table 3). In the second stage, this longer-lived electronic state becomes correlated with the nucleus through the cross-correlation between electron *g*-tensor and anisotropic hyperfine coupling [39]. In the third stage, the microwave Hamiltonian rotates the result into longitudinal electron-nucleus two-spin and three-spin orders, which then efficiently cross-relax (via CSA-HF and HF-HF cross correlations) into nuclear longitudinal magnetisation.

To determine the nature of this effect, we computed laboratory frame BRW relaxation superoperators, both numerically [23,25] and analytically [28], and inspected them term by term alongside the coherent Hamiltonian under the reasonable assumption that the microwave irradiation selectively affects one of the four lines (two exchange-coupled doublets) that are present in the liquid state ESR spectrum of the biradical listed in Table 1. The mechanism that has emerged is summarised in Figure 3 – it appears to be a multi-stage process involving inter-electron exchange coupling and three types of cross-correlations. We have also identified a number of less important processes that operate alongside the mechanism we discuss below; those are too numerous (see the *Mathematica* scripts in the *Supplementary Information*) to discuss in detail. Here we use the product operator formalism [52] to describe the dominant processes.

The three-spin CCDNP mechanism starts with electron longitudinal magnetisation $\hat{E}_Z^{(1,2)}$, which monochromatic microwave irradiation in the presence of weak exchange coupling transforms into:

$$\begin{cases} \hat{E}_Z^{(1)} \\ \hat{E}_Z^{(2)} \end{cases} \xrightarrow{\hat{H}_J, \hat{H}_{MW}} \begin{cases} \hat{E}_+^{(1)} + 2\hat{E}_+^{(1)}\hat{E}_Z^{(2)} \\ \hat{E}_+^{(1)} - 2\hat{E}_+^{(1)}\hat{E}_Z^{(2)} \\ \hat{E}_+^{(2)} + 2\hat{E}_Z^{(1)}\hat{E}_+^{(2)} \\ \hat{E}_+^{(2)} - 2\hat{E}_Z^{(1)}\hat{E}_+^{(2)} \end{cases} \tag{22}$$

The four operators on the right-hand side correspond to the excitation of the four lines that are present in the liquid state ESR spectrum. In a similar way to the TROSY effect in NMR [43], the cross-correlation between the anisotropy of the Zeeman and the dipolar tensor makes the relaxation rates of the four states different [39]; for the example system in Table 1, these are given in Table 3. Automated symbolic processing [28] yields the following relaxation rates for these states [39]:

$$R\left[\hat{E}_+^{(1)} \pm 2\hat{E}_+^{(1)}\hat{E}_Z^{(2)}\right] =$$
$$-\frac{\Delta_{DD}^2}{180}\left[4J(0) + J(\omega_{E2} - \omega_{E1}) + ...\right] - \frac{\Delta_{G1}^2}{45}\left[4J(0) + ...\right] \tag{23}$$
$$-\frac{\Delta_{HF1}^2}{90}\left[2J(0) + 3J(\omega_N) + ...\right] \pm \frac{\aleph_{DD,G1}}{45}\left[4J(0) + ...\right] + ...$$

where the dots indicate the terms that become small (relative to the terms given explicitly) at common NMR fields. The variable sign of the last term is responsible for the differences in the relaxation rates of the four components seen in Table 3. Transverse relaxation times longer than a microsecond are seen, even at high field.

We conclude that modest and technically straightforward microwave field amplitudes would allow this kind of process to take place with acceptable relaxation losses in realistic spin systems. Given the large variation in the relaxation rates of the four components, the presence of the exchange coupling may not be strictly necessary: because the four components would be excited differently by the microwave irradiation even if they overlap – this is illustrated in Figure 2C. In practice, the presence of exchange coupling in biradicals inevitable, but the process described above is not sensitive to its actual amplitude. This is useful because exchange couplings are hard to engineer.

*Table 3. Relaxation rates of the transverse magnetisation operators corresponding to the weak exchange coupling multiplet components of the isotropic EPR spectrum, obtained from a brute-force numerical calculation using Spinach [24] for the spin system described in Table 1. For the analytical expressions, see Equation (23) and the Supplementary Information. Electron Zeeman polarisation at 14.1 Tesla and 298 K is $31.8 \times 10^{-3}$, proton polarisation is $4.83 \times 10^{-5}$.*

| Conditions | $\hat{O}$ | $R[\hat{O}]$, MHz | $\left| \text{Tr}[\hat{\rho}_\infty \hat{O}] \right|$ |
|---|---|---|---|
| $B_0$ = 14.1 T $\Delta\omega_{MW}$ = −0.62 MHz | $\hat{E}_+^{(1)} + 2\hat{E}_+^{(1)}\hat{E}_Z^{(2)}$ | −8.7 | $3.4 \times 10^{-5}$ |
| | $\hat{E}_+^{(1)} - 2\hat{E}_+^{(1)}\hat{E}_Z^{(2)}$ | **−0.49** | **$2.8 \times 10^{-3}$** |
| | $\hat{E}_+^{(2)} + 2\hat{E}_Z^{(1)}\hat{E}_+^{(2)}$ | −74 | $4.2 \times 10^{-6}$ |
| | $\hat{E}_+^{(2)} - 2\hat{E}_Z^{(1)}\hat{E}_+^{(2)}$ | −130 | $1.6 \times 10^{-4}$ |

At the next CCDNP stage, the relaxation superoperator appears to connect the slowly relaxing states from Table 3 to the coherences that involve the nucleus. The corresponding rates are dominated by the cross-correlation terms that, asymptotically, either do not depend on the external magnetic field, or increase with it:

$$R\left[\hat{E}_+^{(1)} \to 2\hat{E}_+^{(1)}\hat{N}_Z\right] = -\frac{\aleph_{G1,HF1}}{45}\left[4J(0) + ...\right]$$

$$R\left[\hat{E}_+^{(1)} \to 4\hat{E}_+^{(1)}\hat{E}_Z^{(2)}\hat{N}_Z\right] = -\frac{\aleph_{DD,HF1}}{90}\left[4J(0) + ...\right] \quad (24)$$

$$R\left[2\hat{E}_+^{(1)}\hat{E}_Z^{(2)} \to 4\hat{E}_+^{(1)}\hat{E}_Z^{(2)}\hat{N}_Z\right] = -\frac{\aleph_{G1,HF1}}{45}\left[4J(0) + ...\right] - ...$$

and likewise for the states with permuted electron indices. Here again, dots indicate the terms that become small (relative to the terms given explicitly) at common NMR fields. The high-field behaviour of these rates is fundamentally better than the longitudinal magnetisation; their numerical values for the system described in Table 1 are given in Table 4.

*Table 4. Cross-correlation rates and steady state populations for the operators featured in Equation (24), obtained from a brute-force numerical calculation using Spinach [24] for the spin system described in Table 1. Electron Zeeman polarisation at 14.1 Tesla and 298 K is $31.8 \times 10^{-3}$, proton polarisation is $4.83 \times 10^{-5}$.*

| Conditions | Process, $\hat{O}_1 \to \hat{O}_2$ | Rate, kHz | $\left|\text{Tr}[\hat{\rho}_\infty \hat{O}_1]\right|$ | $\left|\text{Tr}[\hat{\rho}_\infty \hat{O}_2]\right|$ |
|---|---|---|---|---|
| $B_0$ = 14.1 T $\Delta\omega_{MW}$ = −0.62 MHz | $\hat{E}_+^{(1)} \to 2\hat{E}_+^{(1)}\hat{N}_Z$ | 45.1 | $2.0 \times 10^{-3}$ | $9.6 \times 10^{-5}$ |
| | $\hat{E}_+^{(1)} \to 4\hat{E}_+^{(1)}\hat{E}_Z^{(2)}\hat{N}_Z$ | 42.0 | | $9.6 \times 10^{-5}$ |
| | $2\hat{E}_+^{(1)}\hat{E}_Z^{(2)} \to 4\hat{E}_+^{(1)}\hat{E}_Z^{(2)}\hat{N}_Z$ | 45.1 | $2.0 \times 10^{-3}$ | |

At the final stage of the CCDNP mechanism, the destination states listed in Equation (24) and Table 4 enter another microwave Hamiltonian dynamics loop that connects them to the longitudinal two- and three-spin order that involve the nucleus:

$$\begin{cases} 2\hat{E}_+^{(1,2)}\hat{N}_Z \\ 4\hat{E}_+^{(1,2)}\hat{E}_Z^{(2,1)}\hat{N}_Z \end{cases} \xrightarrow{\hat{H}_{MW}} \begin{cases} 2\hat{E}_Z^{(1,2)}\hat{N}_Z \\ 4\hat{E}_Z^{(1)}\hat{E}_Z^{(2)}\hat{N}_Z \end{cases} \xrightarrow{\hat{R}} \hat{N}_Z \qquad (25)$$

Unlike the electron longitudinal states discussed above, these mixed electron-nuclear states do have a connection to $\hat{N}_Z$ with a favourable magnetic field dependence:

$$R\left[2\hat{E}_Z^{(1)}\hat{N}_Z \to \hat{N}_Z\right] = -\frac{2}{15}\aleph_{CSA,HF1}J(\omega_N)$$

$$R\left[4\hat{E}_Z^{(1)}\hat{E}_Z^{(2)}\hat{N}_Z \to \hat{N}_Z\right] = -\frac{1}{15}\aleph_{HF1,HF2}J(\omega_N) \qquad (26)$$

The numerical values of these rates for the systems described in Table 1 are given in Table 5. The process that involves the longitudinal three-spin order and the HF-HF cross-correlation clearly dominates. This is confirmed by the fact that the nuclear polarisation profile remains unchanged when the nuclear CSA is set to zero (Figure 2D). Thus, there is no requirement in this particular system for the nucleus to be anisotropically shielded. It is possible that the pathways involving HF-CSA cross-correlations could become important in heteronuclear systems that have much larger CSAs.

*Table 5. Cross-correlation rates and steady state populations for the operators featured in Equation (24), obtained from a brute-force numerical calculation using Spinach [24] for the spin system described in Table 1. Electron Zeeman polarisation at 14.1 Tesla and 298 K is 31.8×10⁻³, proton polarisation is 4.83×10⁻⁵.*

| Conditions | Process, $\hat{O}_1 \to \hat{O}_2$ | Rate, Hz | $\left|\text{Tr}\left[\hat{\rho}_\infty \hat{O}_1\right]\right|$ | $\left|\text{Tr}\left[\hat{\rho}_\infty \hat{O}_2\right]\right|$ |
|---|---|---|---|---|
| $B_0$ = 14.1 T $\Delta\omega_{MW}$ = –0.62 MHz | $2\hat{E}_Z^{(1)}\hat{N}_Z \to \hat{N}_Z$ | –6.6 | 5.2×10⁻⁴ | 3.6×10⁻⁴ |
| | $2\hat{E}_Z^{(2)}\hat{N}_Z \to \hat{N}_Z$ | –6.5 | 5.2×10⁻⁴ | |
| | $4\hat{E}_Z^{(1)}\hat{E}_Z^{(2)}\hat{N}_Z \to \hat{N}_Z$ | **–525.1** | **5.3×10⁻⁴** | |

The presence of $J(0)$ terms in the spectral power density parts of Equations (23) and (24) raises the question of the rotational correlation time dependence of the steady state nuclear polarisation. In NMR spectroscopy, $J(0)$ terms have a deleterious effect on transverse relaxation, but numerical simulations (Figures 4 and 5) indicate that they are not a problem here. These simulations suggest that CCDNP is remarkably insensitive to the rotational correlation time (Figure 4A), with an optimum point located close to the correlation times that are expected for common organic biradicals (Figure 5). Cross-correlations are essential here: when *g*-tensor anisotropies are set to zero (Figure 4B), or if one of the electrons is removed (Figure 4C), the nuclear polarisation enhancement disappears.

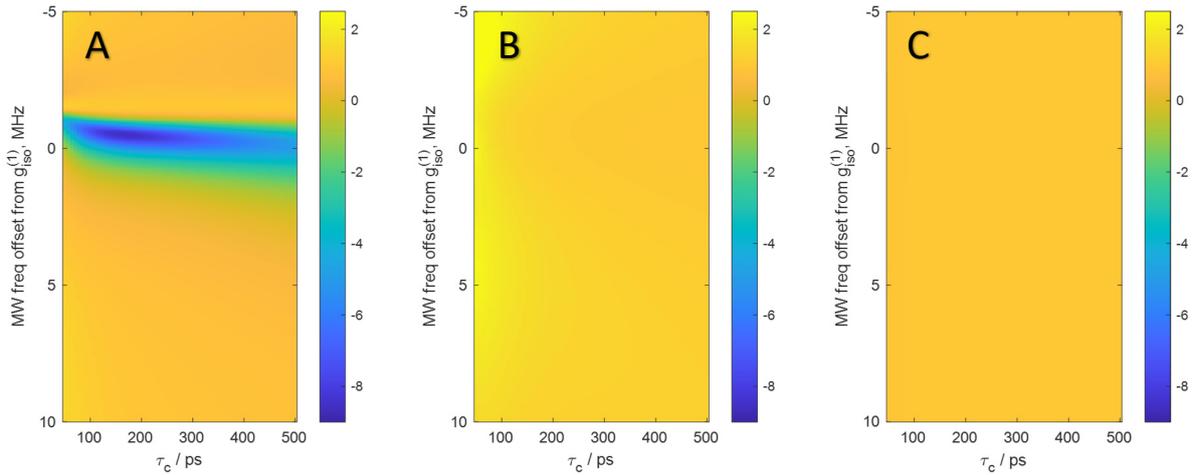

*Figure 4.* Steady state nuclear magnetization relative to the Boltzmann equilibrium level as a function of the microwave offset relative to the Zeeman frequency of electron 1 and rotational correlation time for the spin system specified in Table 1. Simulations were performed at the magnetic field of 14.1 Tesla; the steady state magnetization was normalised to the thermal equilibrium nuclear magnetization at the same field. **(A):** System as specified in Table 1. **(B):** Table 1 but with the g-tensor anisotropy for electron 1 set to zero. **(C):** Table 1 with electron 2 removed.

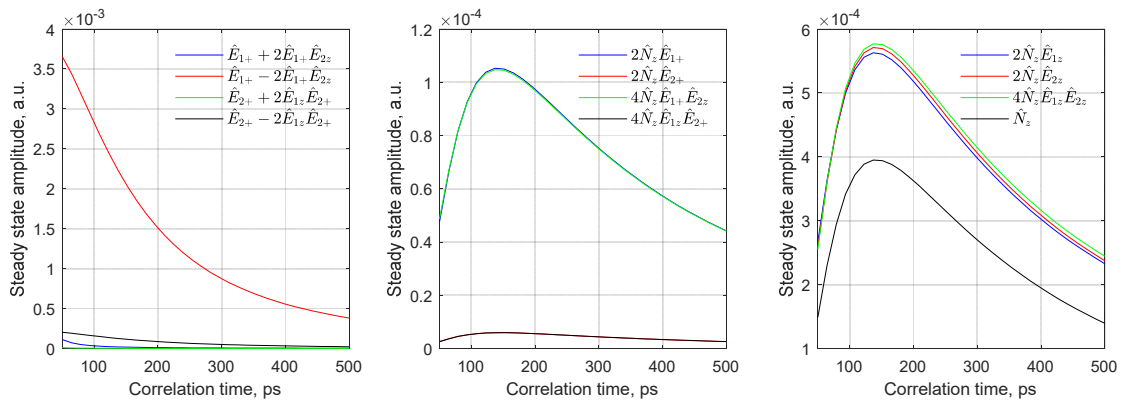

*Figure 5.* Steady state amplitudes (at 14.1 Tesla) of the operators appearing in the mechanism shown in Figure 3 as a function of rotational correlation time. System parameters are listed in Table 1, microwave frequency offset is given in Table 4, external magnetic field is 14.1 Tesla.

## 5. Conclusions and outlook

The content of this paper is a theoretical prediction: detailed simulations indicate that a previously unreported liquid state DNP mechanism exists, with a more favourable high-field behaviour than the nuclear Overhauser effect. This raises the question of whether biradicals with the properties needed for CCDNP can be designed and synthesised. Such biradicals must satisfy the following criteria:

1. Electron Zeeman tensor anisotropy and inter-electron dipolar interaction must be similar at some standard NMR field, such as 14.1 or 18.8 Tesla. Their main axes must be close to collinear, or close to perpendicular, to maximise the TROSY-like effect from the cross-correlation between Zeeman and dipolar tensors.

2. The nucleus must be sufficiently close to at least one of the electrons for the HF-G cross-correlation to be significant. At least one electron-nuclear distance vector must be close to collinear, or close to perpendicular, to the principal axis of the corresponding *g*-tensor, to maximise the HF-G cross-correlation effect.

3. The nucleus must either have a sufficiently large CSA to make the HF-CSA cross-correlation route significant (with the same requirement of parallel or perpendicular alignment of the principal axes), or – preferably – must be sufficiently close to both electrons to make the HF-HF cross-correlation significant (in this case, the optimal location of the nucleus is between the electrons).

4. It is beneficial (though not strictly necessary) to have isotropic *g*-factors of the two electrons sufficiently different and the inter-electron exchange coupling sufficiently strong to resolve four lines in the liquid state ESR spectrum. However, the conformational modulation of the exchange coupling should not dominate electron relaxation.

5. The amplitudes of all interactions and all residence times (with respect to conformational mobility, *etc.*), must be balanced in such a way as to make the overall mechanism generate significant steady-state longitudinal nuclear polarisation. This is possible (Table 1 and two more systems in the SI), but may be hard to engineer.

6. Conditions must exist for the transport of magnetisation from the polarised nucleus to the bulk of the sample. One possibility is a chemical exchange process with residence time in the DNP-active configuration of the order of milliseconds.

This is a tall order, but not an impossible one. Radical pairs showing some aspects of the CCDNP mechanism have been seen in CIDNP systems before [39,40], and the system described in Table 1 is not in any way unusual. The hardest parameter to engineer is the exchange coupling [53], but only a lower bound is needed for it. Experimentally, an NMR/DNP instrument that can scan either the magnetic field and/or the microwave irradiation frequency, would be beneficial; in terms of microwave amplitudes, the requirements are modest. The toughest requirement is probably the need to combine the rigid 2e1n geometry needed for efficient polarisation transfer with the possibility of nuclear magnetisation transfer to the bulk of the sample – either through cross-relaxation or by chemical exchange. Such conditions are known to occur elsewhere in magnetic resonance – for example in relaxation agents and in scalar Overhauser DNP. It remains to be seen if they can be created here.

## Acknowledgements
The authors acknowledge the use of the IRIDIS High Performance Computing Facility, and associated support services at the University of Southampton, in the completion of this work, and the Parallel Computing team at MathWorks for their technical support for the cluster parallel deployment of Matlab. This project was funded by the Weizmann-UK Joint Research Programme, the US National Science Foundation (grants numbers CHE-1808660, DMR-1644779) and the State of Florida.

# High-field solution state DNP using cross-correlations

*Maria Grazia Concilio[1,2], Murari Soundararajan[3], Lucio Frydman[1,3,\*], Ilya Kuprov[2,\*]*

[1]Department of Chemical and Biological Physics, Weizmann Institute of Science, Rehovot, Israel.
[2]School of Chemistry, University of Southampton, Southampton, UK.
[3]National High Magnetic Field Laboratory, Tallahassee, Florida, USA.

This section contains the equivalents of Figures 2-5 and Tables 1-5 in the main text for two other sets of spin Hamiltonian parameters that were found (by a brute-force numerical search) to produce strong steady state nuclear hyperpolarization at high field.

*Table S1.* *The equivalent of Table 1 of the main text for two more sets of spin Hamiltonian parameters. These were found, by a brute-force numerical optimisation of simulated nuclear magnetisation, to have a much larger steady-state DNP amplitude at high field than could be expected from electron-nuclear dipolar cross-relaxation.*

| Parameter | System A | System B |
|---|---|---|
| $^1$H chemical shift tensor eigenvalues, [xx yy zz], ppm | [0 10 20] | [0 10 20] |
| $^1$H chemical shift tensor, ZYZ active Euler angles, rad | [0 0 0] | [0 0 0] |
| Electron 1 *g*-tensor eigenvalues, [xx yy zz] | 1.977873<br>1.977798<br>1.977792 | 1.977800<br>1.977600<br>1.977600 |
| Electron 1 *g*-tensor, ZYZ active Euler angles, rad | [0 0 0] | [−0.180 0.017 0.194] |
| Electron 2 *g*-tensor eigenvalues, [xx yy zz] | 1.977919<br>1.978000<br>1.979000 | 2.006800<br>2.003800<br>2.003800 |
| Electron 2 *g*-tensor, ZYZ active Euler angles, rad | [−0.590 0.100 0.490] | [0.632 0.783 1.086] |
| $^1$H coordinates, [x y z], Angstrom | [0 0 0] | [0 0 0] |
| Electron 1 coordinates, [x y z], Angstrom | [7.0300 0.0187 0.9820] | [6.000 0.030 0.317] |
| Electron 2 coordinates, [x y z], Angstrom | [−7.0300 0.2015 1.0001] | [−6.000 −0.038 0.535] |
| Rotational correlation time, ps | 100 | 100 |
| Electron nutation frequency under MW irradiation, MHz | 1.0 | 1.0 |
| Electron-electron exchange coupling, MHz | 6.2 | 5.0 |

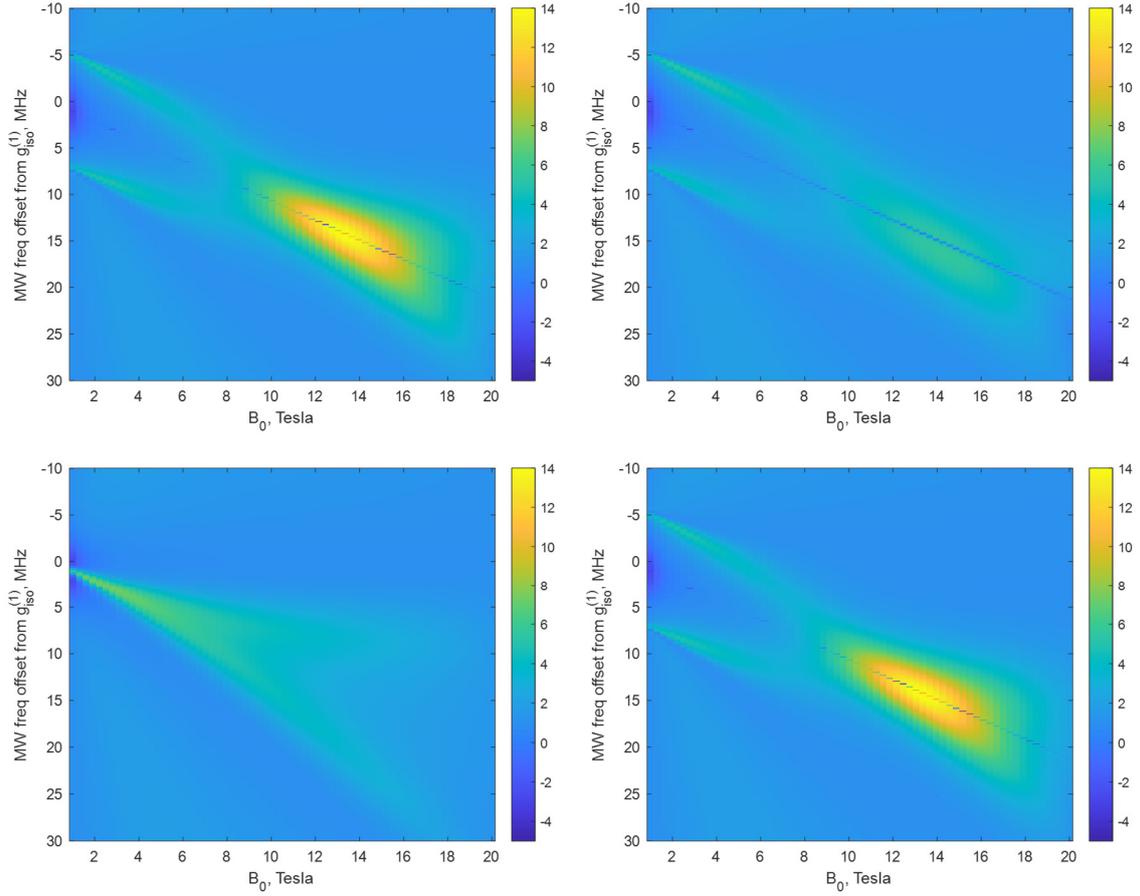

*Figure S1.* The equivalent of Figure 2 of the main text for System A from Table S1: steady state longitudinal nuclear magnetisation (normalised to the thermal equilibrium value) as a function of the static magnetic field and the microwave offset relative to the electron Zeeman frequency. **Top left:** as specified in Table S1. **Top right:** Table S1 with g-tensor anisotropy set to zero on electron 1. **Bottom left:** Table S1 with inter-electron exchange coupling set to zero. **Bottom right:** Table S1 with nuclear chemical shift anisotropy set to zero.

*Table S2.* The equivalent of Table 3 of the main text for the two systems described in Table S1: relaxation rates of the transverse magnetisation operators corresponding to the weak exchange coupling multiplet components of the isotropic EPR spectrum, obtained from a brute-force numerical calculation using Spinach [24]. Electron Zeeman polarisation at 14.1 Tesla and 298 K is $31.8 \times 10^{-3}$, proton polarisation is $4.83 \times 10^{-5}$.

| System | $\hat{O}$ | $R[\hat{O}]$, MHz | $\left\lvert \mathrm{Tr}[\hat{\rho}_\infty \hat{O}] \right\rvert$ |
|---|---|---|---|
| **A**<br>$\Delta\omega_{MW}$ = 15.4 MHz | $\hat{E}_+^{(1)} + 2\hat{E}_+^{(1)}\hat{E}_Z^{(2)}$ | **−0.12** | **$2.3 \times 10^{-4}$** |
| | $\hat{E}_+^{(1)} - 2\hat{E}_+^{(1)}\hat{E}_Z^{(2)}$ | −0.67 | $1.1 \times 10^{-4}$ |
| | $\hat{E}_+^{(2)} + 2\hat{E}_Z^{(1)}\hat{E}_+^{(2)}$ | −0.69 | $1.0 \times 10^{-4}$ |
| | $\hat{E}_+^{(2)} - 2\hat{E}_Z^{(1)}\hat{E}_+^{(2)}$ | **−0.11** | **$2.9 \times 10^{-4}$** |
| **B**<br>$\Delta\omega_{MW}$ = 3.2 MHz | $\hat{E}_+^{(1)} + 2\hat{E}_+^{(1)}\hat{E}_Z^{(2)}$ | **−0.025** | **$2.6 \times 10^{-3}$** |
| | $\hat{E}_+^{(1)} - 2\hat{E}_+^{(1)}\hat{E}_Z^{(2)}$ | −2.5 | $2.6 \times 10^{-4}$ |
| | $\hat{E}_+^{(2)} + 2\hat{E}_Z^{(1)}\hat{E}_+^{(2)}$ | −131 | $8.1 \times 10^{-7}$ |
| | $\hat{E}_+^{(2)} - 2\hat{E}_Z^{(1)}\hat{E}_+^{(2)}$ | −116 | $4.4 \times 10^{-7}$ |

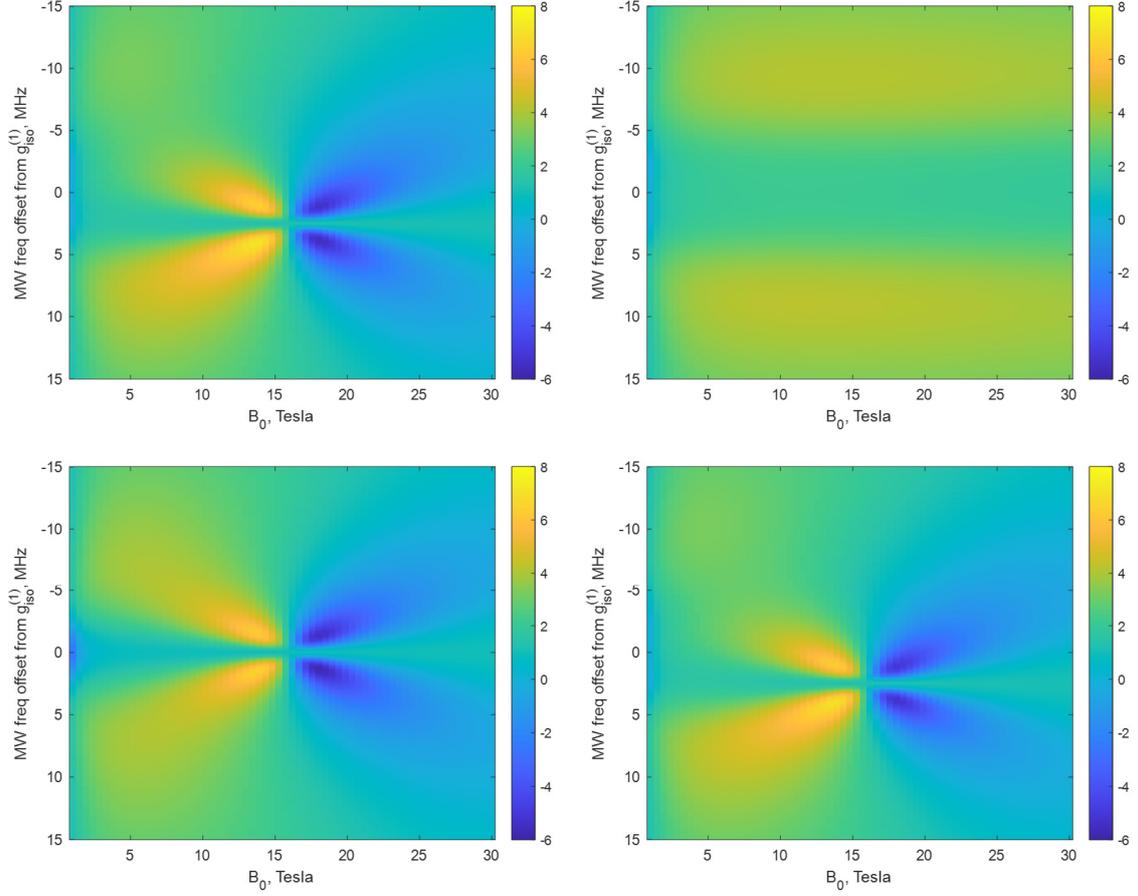

*Figure S2. The equivalent of Figure 2 of the main text for System B from Table S1: steady state longitudinal nuclear magnet-isation (normalised to the thermal equilibrium value) as a function of the static magnetic field and the microwave offset relative to the electron Zeeman frequency. **Top left:** as specified in Table S1. **Top right:** Table S1 with g-tensor anisotropy set to zero on electron 1. **Bottom left:** Table S1 with inter-electron exchange coupling set to zero. **Bottom right:** Table S1 with nuclear chemical shift anisotropy set to zero.*

*Table S3. The equivalent of Table 4 of the main text for the two systems described in Table S1: cross-correlation rates and steady state populations for the operators featured in Equation (24), obtained from a brute-force numerical calculation using Spinach [24]. Electron Zeeman polarisation at 14.1 Tesla and 298 K is 31.8×10⁻³, proton polarisation is 4.83×10⁻⁵.*

| System | Process, $\hat{O}_1 \to \hat{O}_2$ | Rate, kHz | $\left\| \mathrm{Tr}\left[\hat{\rho}_\infty \hat{O}_1\right] \right\|$ | $\left\| \mathrm{Tr}\left[\hat{\rho}_\infty \hat{O}_2\right] \right\|$ |
|---|---|---|---|---|
| **A** $\Delta\omega_{\mathrm{MW}}$ = 15.4 MHz | $\hat{E}_+^{(1)} \to 2\hat{E}_+^{(1)}\hat{N}_Z$ | −3.4 | 2.4×10⁻⁴ | 2.8×10⁻⁶ |
| | $\hat{E}_+^{(1)} \to 4\hat{E}_+^{(1)}\hat{E}_Z^{(2)}\hat{N}_Z$ | 6.1 | | 6.9×10⁻⁶ |
| | $2\hat{E}_+^{(1)}\hat{E}_Z^{(2)} \to 4\hat{E}_+^{(1)}\hat{E}_Z^{(2)}\hat{N}_Z$ | −3.4 | 8.3×10⁻⁵ | |
| **B** $\Delta\omega_{\mathrm{MW}}$ = 3.2 MHz | $\hat{E}_+^{(1)} \to 2\hat{E}_+^{(1)}\hat{N}_Z$ | −14.8 | 2.0×10⁻³ | 6.5×10⁻⁵ |
| | $\hat{E}_+^{(1)} \to 4\hat{E}_+^{(1)}\hat{E}_Z^{(2)}\hat{N}_Z$ | −16.8 | | 6.9×10⁻⁵ |
| | $2\hat{E}_+^{(1)}\hat{E}_Z^{(2)} \to 4\hat{E}_+^{(1)}\hat{E}_Z^{(2)}\hat{N}_Z$ | −14.8 | 1.6×10⁻³ | |

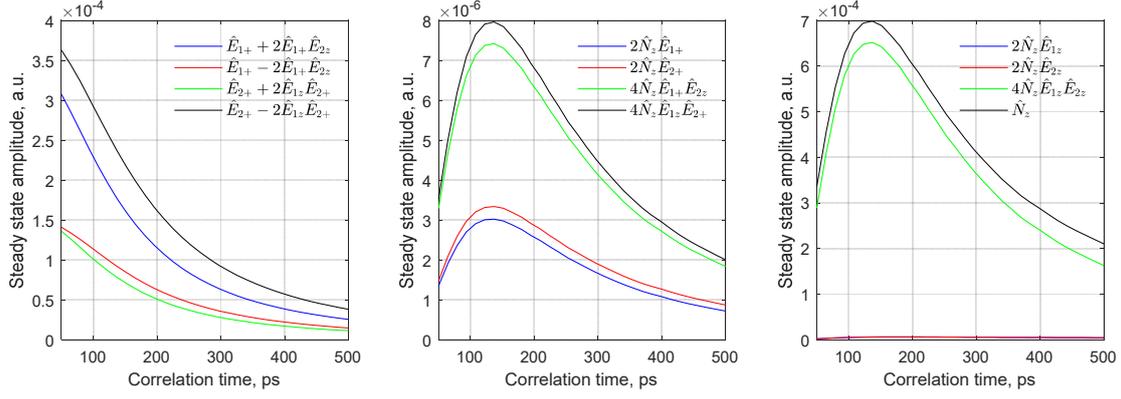

***Figure S3.*** *The equivalent of Figure 5 of the main text for System A from Table S1: steady state amplitudes (at 14.1 Tesla) of the operators appearing in the mechanism shown in Figure 3 as a function of rotational correlation time. Microwave frequency offsets are is given in Table S3; external magnetic field is 14.1 Tesla.*

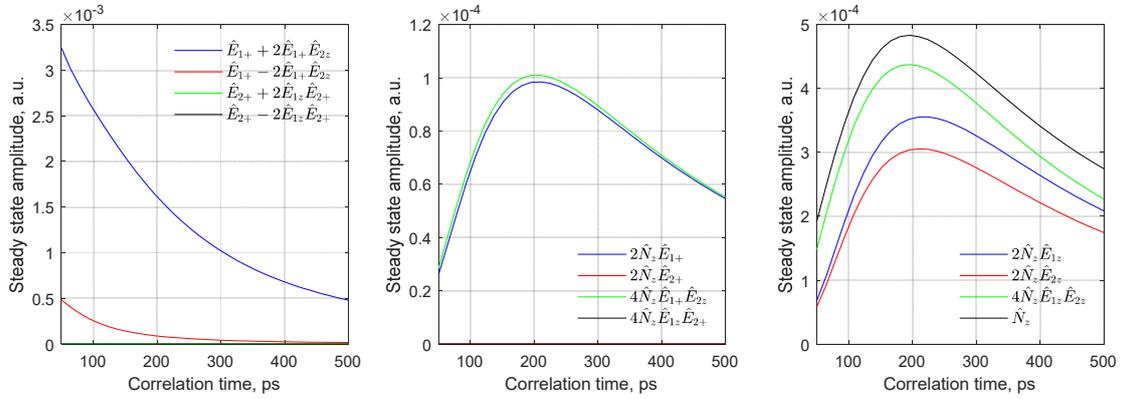

***Figure S4.*** *The equivalent of Figure 5 of the main text for System B from Table S1: steady state amplitudes (at 14.1 Tesla) of the operators appearing in the mechanism shown in Figure 3 as a function of rotational correlation time. Microwave frequency offsets are is given in Table S3; external magnetic field is 14.1 Tesla.*

***Table S4.*** *The equivalent of Table 5 of the main text for the spin systems from Table S1: cross-correlation rates and steady state populations for the operators featured in Equation (24), obtained from a brute-force numerical calculation using Spinach [24]. Electron Zeeman polarisation at 14.1 Tesla and 298 K is $31.8\times10^{-3}$, proton polarisation is $4.83\times10^{-5}$.*

| System | Process, $\hat{O}_1 \to \hat{O}_2$ | Rate, Hz | $\left|\text{Tr}\left[\hat{\rho}_\infty \hat{O}_1\right]\right|$ | $\left|\text{Tr}\left[\hat{\rho}_\infty \hat{O}_2\right]\right|$ |
|---|---|---|---|---|
| **A** $\Delta\omega_{MW}$ = 15.4 MHz | $2\hat{E}_Z^{(1)}\hat{N}_Z \to \hat{N}_Z$ | −2.6 | $5.3\times10^{-6}$ | $6.5\times10^{-4}$ |
| | $2\hat{E}_Z^{(2)}\hat{N}_Z \to \hat{N}_Z$ | −2.6 | $5.2\times10^{-6}$ | |
| | $4\hat{E}_Z^{(1)}\hat{E}_Z^{(2)}\hat{N}_Z \to \hat{N}_Z$ | **−98.5** | **$6.0\times10^{-4}$** | |
| **B** $\Delta\omega_{MW}$ = 3.2 MHz | $2\hat{E}_Z^{(1)}\hat{N}_Z \to \hat{N}_Z$ | −4.4 | $2.1\times10^{-4}$ | $3.6\times10^{-4}$ |
| | $2\hat{E}_Z^{(2)}\hat{N}_Z \to \hat{N}_Z$ | −4.4 | $1.8\times10^{-4}$ | |
| | $4\hat{E}_Z^{(1)}\hat{E}_Z^{(2)}\hat{N}_Z \to \hat{N}_Z$ | **−262.2** | **$3.2\times10^{-4}$** | |

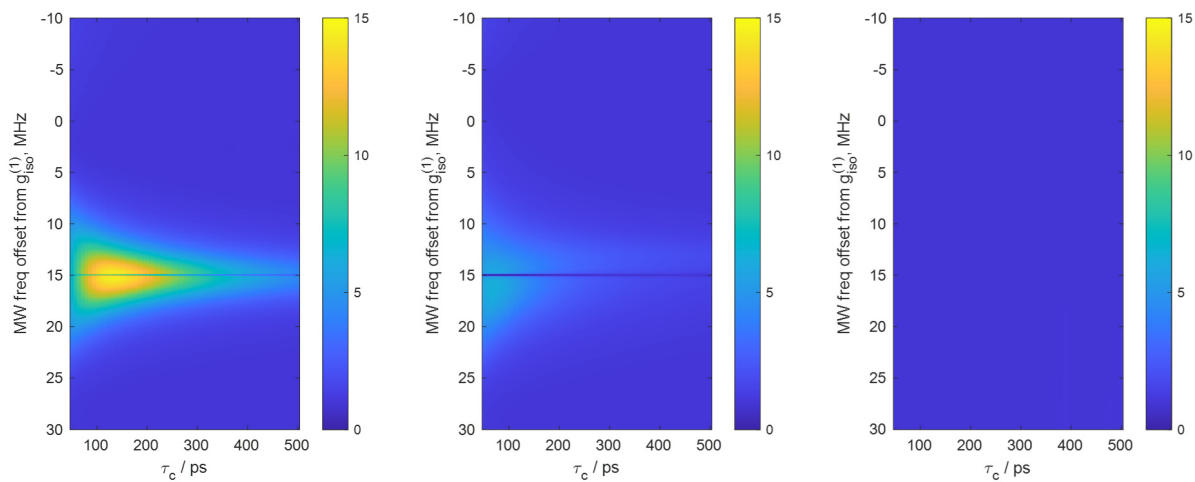

*Figure S5.* The equivalent of Figure 4 of the main text for System A from Table S1: steady state nuclear magnetization relative to the Boltzmann equilibrium level as a function of the microwave offset relative to the Zeeman frequency of electron 1 and rotational correlation time. Simulations were performed at the magnetic field of 14.1 Tesla; the steady state magnetization was normalised to the thermal equilibrium nuclear magnetization at the same field. **Left:** system as specified in Table S1. **Middle:** Table S1 with g-tensor anisotropy for electron 1 set to zero. **Right:** Table S1 with electron 2 removed.

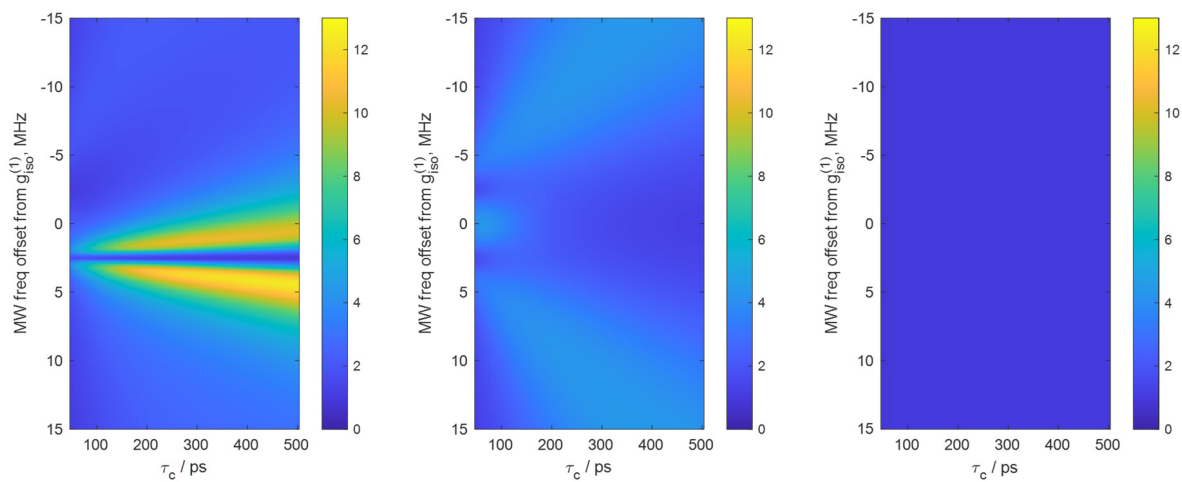

*Figure S6.* The equivalent of Figure 4 of the main text for System B from Table S1: steady state nuclear magnetization relative to the Boltzmann equilibrium level as a function of the microwave offset relative to the Zeeman frequency of electron 1 and rotational correlation time. Simulations were performed at the magnetic field of 14.1 Tesla; the steady state magnetization was normalised to the thermal equilibrium nuclear magnetization at the same field. **Left:** system as specified in Table S1. **Middle:** Table S1 with g-tensor anisotropy for electron 1 set to zero. **Right:** Table S1 with electron 2 removed.